\documentclass[preprint]{imsart}
\pubyear{2022}
\arxiv{}
\firstpage{1}
\lastpage{1}

\usepackage{amsthm}
\usepackage{amsmath}
\usepackage{amsfonts}
\usepackage{natbib}
\usepackage[colorlinks,citecolor=blue,urlcolor=blue,filecolor=blue,backref=page]{hyperref}
\usepackage{graphicx}
\usepackage{comment}
\usepackage{subfig}

\startlocaldefs

\usepackage[export]{adjustbox}
\newcommand{\N}[1]{\mathrm{N}\!\left(#1\right)}
\newcommand{\logit}[1]{\mathrm{logit}\left(#1\right)}
\newcommand{\cI}{\mathcal{I}}
\newcommand{\bmu}{\boldsymbol\mu}
\newcommand{\obs}{\mathrm{obs}}
\newcommand{\miss}{\mathrm{miss}}
\newcommand{\new}{\mathrm{new}}
\newcommand{\ret}{\mathrm{ret}}
\endlocaldefs

\begin{document}


\begin{frontmatter}
\title{Detecting changes in dynamic social networks using multiply-labeled movement data}
\runtitle{Dynamic social networks using multiply-labeled movement data}

\begin{aug}
\author{\fnms{Zaineb L.} \snm{Boulil}\thanksref{addr1}},
\author{\fnms{John W.} \snm{Durban}\thanksref{addr2}},
\author{\fnms{Holly} \snm{Fearnbach}\thanksref{addr3}},
\author{\fnms{Trevor W.} \snm{Joyce}\thanksref{addr4}},
\author{\fnms{Samantha G. M.} \snm{Leander}\thanksref{addr2}},
\and
\author{\fnms{Henry R.} \snm{Scharf}\thanksref{addr1}\ead[label=e1]{hscharf@sdsu.edu}}

\runauthor{Z. Boulil et al.}

\address[addr1]{San Diego State University, 
  Department of Mathematics and Statistics, 
  5500 Campanile Dr., 
  San Diego, CA 92182, U.S.A.}
\address[addr2]{Southall Environmental Associates, Inc., 
  9099 Soquel Drive, 
  Aptos, CA 95003, U.S.A.}
\address[addr3]{SR3 SeaLife Response, Rehabilitation and Research, 
  2003 S. 216th St. \#98811, 
  Des Moines, WA, 98198, U.S.A.}
\address[addr4]{Environmental Assessment Services, 
  350 Hills St., Suite 112, 
  Richland, WA 99354, U.S.A.}


\end{aug}

\begin{abstract}
The social structure of an animal population can often influence movement and inform researchers on a species’ behavioral tendencies. Animal social networks can be studied through movement data; however, modern sources of data can have identification issues that result in multiply-labeled individuals. Since all available social movement models rely on unique labels, we extend an existing Bayesian hierarchical movement model in a way that makes use of a latent  social network and accommodates multiply-labeled movement data (MLMD). We apply our model to drone-measured movement data from Risso's dolphins (\textit{Grampus griseus}) and estimate the effects of sonar exposure on the dolphins’ social structure. Our proposed framework can be applied to MLMD for various social movement applications.
\end{abstract}


\end{frontmatter}


\section{Introduction}  \label{sec:intro}
For many species, analyzing social networks is necessary for understanding animal behavior \citep[e.g.,][]{couzin2005effective}. The underlying social network of an animal population can strongly influence movement \citep{scharf2016dynamic, scharf2018processes} and provide researchers with  information on the behavioral characteristics of a species. Such information is beneficial to various ecological applications, including  determining the effects of anthropogenic activities, landscape fragmentation, or pollutants on the collective behavior of an animal population.

Many marine mammals, including Risso's dolphins (\textit{Grampus griseus}), are highly social with close-knit relationships \citep{lusseau2004role, hartman2008stratified, weiss2021diversity, whitehead2021cultural}. With such an affinity for social interactions, they may be expected to exhibit a collective behavioral response when exposed to changes in their environment. For example, \textit{G. Griseus} are commonly exposed to mid-frequency active sonar due to their high abundance in near-coastal navy training areas \citep{caretta2019stock, rice2019acoustic}. Nonetheless, there is a sparsity of data on how dolphins respond to sonar and uncertainty about the consequences of exposure  \citep{durban2022integrating}. Thus, our motivating application is studying \textit{G. Griseus} exposed to sonar off Catalina Island in Southern California in a controlled exposure experiment (CEE), designed to fill these key data gaps. 

Currently available social movement models that account for social interactions between individuals require unique identification of animals and assume a closed, fully observed population throughout the study period \citep[e.g.,][]{langrock2014modelling, scharf2016dynamic, scharf2018processes, niu2020modelling, scharf2020animal, milner2021modelling}. Movement data of terrestrial and marine animals have been conventionally obtained by tracking individuals singularly, often with telemetry tags. However, drones now offer the ability to track multiple individuals simultaneously \citep{durban2022integrating}. While this new source of movement data can achieve a high spatial precision \citep{dawsonphotgrammetry2017, durban2015hexacopter, durban2022integrating}, it sometimes introduces complex labeling issues due to the observation process. When animals that cannot be uniquely identified disappear out of a drone's active field of view for sustained intervals of time, they are typically assigned a new label upon reappearance. The lack of identification leads to problematic ``multi-labeling'' as well as varying numbers of animals in view at a given time point. Thus, new methods are needed that allow for social network inference from multiply-labeled movement data (MLMD) such as those obtained via drones. We extend existing methodology for animal movement to estimate the effects of sonar exposure on an unobserved social network of dolphins through the use of drone-measured movement data. Our modeling approach allows us to infer social connections between dolphins directly from the drone-measured movement data.
 
We adopt a Bayesian approach through the implementation of a discrete-time continuous-space Gaussian Markov Random Field \citep[GMRF;][]{rue2005gaussian} with an underlying dynamic social network. The two main behavioral components motivating the model are attraction and alignment, both of which are directly related to social structure. Attraction represents an individual’s inclination toward the mean position of connected individuals, while alignment represents an individual’s tendency to move in parallel with the trajectories of connected individuals \citep{bode2012distinguishing}. The model also allows for repulsive and/or anti-aligning behavior by allowing negative values of relevant parameters, however, such behaviors are unlikely to occur in our motivating application. Thus, we limit our framework to modeling the positive behaviors of attraction and alignment. As social structure is expected to play a role in the movement of animals in relation to one another, the attraction and alignment mechanisms are motivating elements influencing animal movement. Additionally, the model captures the overall stability of the social network and measures the density of social connections as it varies with time. 

We incorporate generalized linear models (GLMs) on certain parameters of interest as an extension to the model introduced in \citet{scharf2016dynamic}. The GLMs provide information on the direct impact of external sources on the behavioral characteristics of a population, which allow for assessing the effect of environmental covariates on social structure. By including GLMs in the modeling framework, we provide a deeper understanding of how external sources might inﬂuence the collective movement of a population.

While the closely related model from \citet{scharf2016dynamic} provides a tractable likelihood for individuals with unique labels, the multiply-labeled nature of MLMD makes calculating the likelihood infeasible. To accommodate the intractability of the exact likelihood, we develop an implementation scheme based on a proxy likelihood, which we evaluate through a simulation study.

We detail our inference of social networks when analyzing MLMD in Section \ref{sec:methods} and \ref{sec:imp} and validate our approach using an approximate likelihood through a simulation study in Section \ref{sec:sim}. We apply our methodology to drone-measured movement data following many Risso's dolphins simultaneously as they are exposed to sonar in Section \ref{sec:app}. We found that during periods of sonar exposure, dolphins are more likely to exhibit an inclination towards the mean position of connected individuals, less likely to move in parallel with connected individuals, and tend to have fewer social connections. We close with potential for future directions in Section \ref{sec:dis} including applying our model to subsets of a population when analyzing data sets with a large number of individuals. 

\section{Methods} \label{sec:methods} 
We build upon a discrete-time continuous-space hierarchical GMRF model implemented in previous social movement research \citep{scharf2016dynamic}. We select a GMRF as it models dependence between neighbors in a graphical sense, a phenomenon that translates naturally when assessing social movement where connected animals “neighbor” one another. Additionally, the discrete-time approach of the GMRF model aligns well with our motivating application as the data consist of regularly-observed positions.

We begin with a brief review of the model for social movement from \cite{scharf2016dynamic}, then extend the model to allow for time-varying parameters. Let $\bmu_i(t)$ denote the column vector representing the position (e.g., longitude and latitude) of individual $i$ at time $t$, and let $\bmu(t) = \left(\bmu'_1(t), \dots, \bmu'_J(t)\right)'$ denote the concatenation of the positions of all $J$ individuals at time $t$. In addition, let $\mathbf{W} = \left\{\mathbf{W}(t); t = 1, \dots T \right\}$ denote a set of time-indexed social network adjacency matrices such that elements $w_{ij}(t)$ are binary variables equal to 1 when $i$ and $j$ share a social connection at time $t$, and 0 otherwise. \cite{scharf2016dynamic} specify a Gaussian joint model for the positions of all individuals conditioned on the positions at the previous time point as
\begin{align}
  \bmu(t)|\bmu(t - 1), \mathbf{W}(t - 1), \alpha , \beta , \sigma^{2} \sim
  \N{\bmu(t - 1) + \beta \left(\bar\bmu(t - 1) - \bmu(t - 1)\right), 
    \; \sigma^2 \mathbf{Q}^{-1}(t - 1)}, \label{eqn:mu_t}
 \end{align}
with precision matrix,
\begin{align}
  Q_{ij}(t) &\equiv \begin{cases}
    -\alpha w_{ij}(t), \quad & j \neq i \\
    w_{i+}(t), & j = i,
  \end{cases}
  \label{eqn:q}
\end{align}
where $w_{ii}(t) = 1$ and $w_{i+}(t) = \sum_{j=1}^J w_{ij}(t)$ is the number of connections each individual has, or the size of its ego-network.

The vector $\bar\bmu(t) = \left( \bar\bmu'_1(t), \dots, \bar\bmu'_J(t)\right)' $ is a collection of sub-vectors for each individual representing the mean position of all socially-connected individuals, such that
\begin{align}
  \bar\bmu_i(t) = \sum_{j = 1:J}\frac{w_{ij}(t)}{w_{i+}(t)}\bmu_j(t). 
\end{align}
Thus, the expected displacement of each individual at time $t$ is given by a vector pointing from its position at time $t-1$ towards its neighbors' previous positions, scaled by $\beta$. The joint dependence across all individuals' displacements is also related to the social connections, and the strength of dependence is controlled by $\alpha$.

\cite{scharf2016dynamic} show how the parameters $\beta$ and $\alpha$ can be interpreted via two behavioral mechanisms: attraction and alignment. The strength of the attraction component is controlled by $\beta$, and describes the tendency of an individual towards the mean position of its connected individuals. Values of $\beta$ near 0 correspond to no attraction while values near 1 correspond to complete attraction where, on average, an individual closes the full distance between itself and the center of its connected individuals over each time step. 

The strength of the alignment term is controlled by $\alpha$ and describes the tendency of an individual to move parallel to the expected displacement of its connected individuals from time $t-1$ to time $t$. Values of $\alpha$ near 0 correspond to no alignment while values near 1 correspond to perfect alignment where an individual moves in parallel with its connected individuals. The parameter $\sigma$ scales the process so that step lengths are consistent with the appropriate spatial dimensions.

In \cite{scharf2016dynamic}, the value of $w_{ii}(t)$ was taken to be $0$ and $w_{i+}^c(t) = \mathrm{max}\left\{w_{i+}(t), c\right\}, c > 0$ was defined in lieu of $w_{i+}(t)$ to cover the possibility of zero social connections. In the present work, we take $w_{ii}(t) = 1$, which obviates $c$ and only changes the interpretation of a single parameter, $\beta$, in the model. Under the modification that includes self-edges, the same magnitude of $\beta$ is associated with a slightly weaker attraction effect than in \cite{scharf2016dynamic}, although the interpretation of the sign is unchanged.

\cite{scharf2016dynamic} formulate the dynamic binary network, $\mathbf{W}$, from a pairwise-independent set of stationary Markov processes initialized as $w_{ij}(1) \sim \mathrm{Bern}(p_1)$, and defined for $t > 1$ as 
\begin{align}
  w_{ij}(t)|w_{ij}(t - 1) &\sim
  \begin{cases}
    \mathrm{Bern}(p_{1|0}), \; &w_{ij}(t - 1) = 0\\
    \mathrm{Bern}(p_{1|1}), &w_{ij}(t - 1) = 1,
  \end{cases} \label{eqn:wt}
\end{align}
where $p_{1|0} = (1 - \phi) p_{1}$ and $p_{1|1} = 1 - (1 - \phi)(1 - p_{1})$.

The parameter $p_{1} \in [0, 1]$ controls the connectivity between individuals within the latent network process and represents the marginal probability of a connection between any two individuals during the observation period. The parameter $\phi \in [0, 1]$ controls the overall stability of the network. Higher values of $\phi$ indicate the connections within the study population are stable and vary little with time, whereas lower values mean individuals are less likely to maintain connections for longer periods of time. Values of $\phi$ = 0 and $\phi$ = 1 correspond to complete temporal independence, and complete temporal dependence (i.e., a completely static network with $w_{ij}(t) = w_{ij}(1), \forall i, j, t$), respectively.

Animal movement can be affected by a variety of anthropogenic and environmental covariates. To determine the impact of external sources on the behavioral characteristics of a species, we incorporate generalized linear models on a subset of parameters that give rise to the position process and underlying social network, thereby generalizing the static parameters controlling the strength of attraction, alignment, and the overall connectivity of the population in \cite{scharf2016dynamic} to allow them to vary with time in response to measured covariates. We select logit links for each GLM as we constrain the parameters, $\alpha(t)$ $\beta(t)$ and $p_{1}(t)$, to values ranging between 0 and 1, such that
\begin{align}
\logit{\alpha(t)} = \mathbf{x}^\prime(t) \boldsymbol\delta_\alpha, \quad
\logit{\beta(t)} = \mathbf{x}^\prime(t) \boldsymbol\delta_\beta, \quad
\logit{p_{1}(t)} = \mathbf{x}^\prime(t) \boldsymbol\delta_{p_{1}},
\end{align}
where $\mathbf{x}(t)$ are the time-indexed covariates of interest.

Depending on the scientific goals of an analysis, we note that the same GLM structure could also be utilized for additional dynamic variables (e.g., $\phi$). Similarly, the logit link functions are specifically chosen in the context of our application, however alternative link functions may be used. For example, modeling repulsion instead of attraction and/or anti-alignment instead of alignment would require negative values of $\beta(t)$ and $\alpha(t)$, which would in turn necessitate an alternative to the logit link function. The GLM framework also has the capability to include both fixed and random effects and a variety of time-varying environmental covariates, which can be grouped or specific to individual parameters. 

We take advantage of the linear version of the GMRF introduced in \cite{scharf2016dynamic} and write the model for movement as 
\begin{align}
  \bmu(t)|\bmu(t - 1), \boldsymbol\theta, \mathbf{W}(t - 1) \sim
  \N{\mathbf{A}(t - 1) \bmu(t - 1), \; \sigma^2 \mathbf{Q}^{-1}(t - 1)},& \label{eqn:mu_t2}
\end{align}
where $\boldsymbol\theta = \left(\boldsymbol\delta'_\alpha , \boldsymbol\delta'_\beta, \sigma^{2}\right)'$ is a vector of model parameters, 
\begin{align}
  \mathbf{A}(t) &= (1 - \beta(t))\mathbf{I} + \beta(t) \overline{\mathbf{W}}(t), \label{eqn:A_matrix} \\
  \overline{w}_{ij}(t) &= w_{ij} / w_{i+}(t),
\end{align}
and the precision matrix, $\mathbf{Q}(t)$, is defined analogously to \eqref{eqn:q} now with $\alpha(t)$. 

\section{Implementation} \label{sec:imp}
The hierarchical model described in Section~\ref{sec:methods} presumes fully-labeled, uninterrupted observations of all individuals throughout the study period. In our application, the drone-based data represent a partially censored subset of the full trajectories. Not all individuals are captured in the camera's field of view at all times, and individuals are not always uniquely identifiable visually, leading to multiple labels for the same individuals. Ideally, the missing positions and labels could be treated as latent random variables that would then be marginalized over to yield a likelihood for the observed data, and inference could proceed using standard methods. However, the complexity and dimensionality of the missing variables renders such an approach infeasible (see Supplementary Materials). We propose a proxy for the true, marginalized likelihood function based on a product of component likelihoods. We evaluate the consistency of our approach using a comprehensive simulation study and find that it performs adequately well for the primary parameters of interest in our motivating application.

For the case of zero censoring, we let $\bmu = \{ \bmu_i(t): i = 1, \dots, n; t = 1, \dots, T \}$ denote the concatenation of positions for all individuals at all times and refer to it as the `complete' data. The conditional likelihood for the vector of parameters, $\boldsymbol\theta = (\boldsymbol\delta_\alpha', \boldsymbol\delta_\beta', \sigma^2)'$, given a dynamic network, $\mathbf{W}$, is
\begin{align}
L(\boldsymbol\theta | \mathbf{W}, \bmu) = \prod_{t = 2}^T \N{\bmu(t); \mathbf{A}(t)\bmu(t - 1),
  \sigma^2\mathbf{Q}^{-1}(t - 1)}, \label{eqn:complete_likelihood}
\end{align}
where $\N{\mathbf{x} ; \mathbf{m}, \boldsymbol\Sigma}$ denotes the multivariate Gaussian density function for a random vector $\mathbf{x}$ with mean $\mathbf{m}$ and covariance matrix $\boldsymbol\Sigma$. The full-conditional distribution for $\boldsymbol\theta$ is proportional to the product of $L(\boldsymbol\theta | \mathbf{W}, \bmu)$ and the prior distribution for $\boldsymbol\theta$, and is therefore required for the application of conventional MCMC algorithms.

Were the correct individual labels available for all observed positions, the exact conditional likelihood for the observed (non-missing) data could be evaluated directly by integrating over the unobserved locations and would require at most the same computing time as the complete likelihood depending on the pattern of missingness. Alternatively, the missing observations could be treated as random variables to be updated in an MCMC algorithm. For the case of MLMD, however, the correct labels are unknown.

\subsection{Approximate likelihood}\label{sec:proxy_likelihood}

One might consider developing a model for MLMD incorporating random permutations and censoring as part of the observation process. However, integrating over such random permutations quickly leads to an intractable problem as the number of affected time points grows (see Supplementary Materials). Attempting to update the random permutations, as well as the missing positions, in an augmented MCMC algorithm would involve huge numbers of updates of highly dependent variables. Such an approach is very likely to suffer from inefficient and computationally expensive algorithms. To circumvent these analytical and computational obstacles, we pursue a simpler, albeit approximate, alternative. 

We propose a proxy for the marginalized likelihood motivated by taking the perspective of a naive observer who is unaware of individuals outside the field of view. Briefly, our proposed proxy likelihood is constructed by replacing each multivariate Gaussian density in \eqref{eqn:complete_likelihood} with an analogous multivariate Gaussian density that corresponds to the instantaneous assumption that the entire population is limited to those individuals observed at both the previous and current time points (i.e., the ``returners'').
To make the proposed proxy likelihood mathematically precise, we first establish some useful notation. Let $\cI_\obs(t)$ be an index for the positions of all observed individuals, which may vary in size, $n_\obs(t) = |\cI_\obs(t)|$, and let $\bmu_\obs(t) = \{ \bmu_i(t): i \in \cI_\obs(t)\}$ denote the positions of all observed individuals at time $t$, with $\bmu_\obs = \{\bmu_\obs(t): t = 1, \dots, T \}$ denoting the concatenation of all observed positions throughout the entire study period. In addition, let $\cI_\ret(t) = \cI_\obs(t) \cap \cI_\obs(t - 1)$ index those individuals observed contiguously from time $t-1$ to $t$ (the ``returners'') of size $|\cI_\ret(t)| = n_\ret(t) \leq n_\obs(t)$, and define $\bmu_\ret(t)$ and $\bmu_\ret$ analogously to the observed positions. Define the proxy likelihood as

\begin{align}
\tilde{L}(\boldsymbol\theta | \mathbf{W}_\ret, \bmu_\ret) &= \prod_{t = 2}^T \N{\bmu_\ret(t); \mathbf{A}_\ret(t)\bmu_\ret(t - 1), \sigma^{2}\mathbf{Q}_\ret^{-1}(t)},
\end{align}
where $\mathbf{W}_\ret(t)$ is the sub-matrix of $\mathbf{W}(t)$ whose rows and columns correspond to $\cI_\ret(t)$, $\mathbf{W}_\ret = \{\mathbf{W}_\ret(t): t = 1, \dots, T\}$, and $\mathbf{A}_\ret(t)$ and $\mathbf{Q}_\ret(t)$ are constructed from $\mathbf{W}_\ret(t)$ in lieu of $\mathbf{W}(t)$.

Although the proxy likelihood simplifies to the complete likelihood for the special case where no censoring takes place, in general, the proxy likelihood may be very different from the true likelihood $L(\boldsymbol\theta | \mathbf{W}, \bmu_\obs)$, obtained by marginalizing over all unobserved positions and the unknown censoring/labeling process. Nevertheless, $\tilde{L}(\boldsymbol\theta |\mathbf{W}_\ret, \bmu_\ret)$ is an accessible alternative that still encodes the mechanisms of attraction and alignment conditioned on an underlying dynamic social network, and is therefore worth consideration as a potentially useful proxy. We investigate the feasibility of substituting the proposed proxy likelihood in place of the marginalized likelihood for Bayesian inference using a comprehensive simulation study that considers a range of possible parameter values and censoring mechanisms. 

\subsection{Approximate distribution for dynamic network}\label{sec:proxy_network}

An initial goal in our application is to infer global characteristics of the latent dynamic social network that drives movement. For uncensored movement data, it is straightforward to obtain samples from the posterior distribution of the network using Gibbs updates for each connection $w_{ij}(t)$ within an MCMC algorithm. Because of the Markov structure in both the position and network process, in the absence of censoring, the full-conditional densities for each $w_{ij}(t)$ depend only on the densities of $\bmu(t) | \boldsymbol\theta$, $w_{ij}(t) | w_{ij}(t - 1)$, and $w_{ij}(t + 1) | w_{ij}(t)$. However, censoring and multi-labeling require modifications because the indices $i$ and $j$ are not consistent in time. Analogous to the proxy likelihood in Section~\ref{sec:proxy_likelihood}, we propose substituting the Markov process for the dynamic social network based on complete data with one restricted to observed individuals.

Let $\cI_\new(t) = \cI_\obs(t) \setminus \cI_\ret(t)$ be the index of individuals observed at time $t$, but not at time $t - 1$ (the ``newcomers'') such that $\cI_\new(t)$ and $\cI_\ret(t)$ partition the index of observed individuals at time $t$. For $t > 1$ and $i, j \in \cI_\obs(t)$, define the proxy conditional probability mass function for each edge variable as
\begin{align}
  \tilde{p}\left(w_{ij}(t) | w_{ij}(t - 1)\right) = 
    \begin{cases}
      p_{1|0}(t)^{w_{ij}(t)}(1 - p_{1|0}(t))^{1 - w_{ij}(t)}, & w_{ij}(t - 1) = 0, \quad i, j \in \cI_\ret(t) \\
      p_{1|1}(t)^{w_{ij}(t)}(1 - p_{1|1}(t))^{1 - w_{ij}(t)}, & w_{ij}(t - 1) = 1, \quad i, j \in \cI_\ret(t) \\
      p_1(t)^{w_{ij}(t)}(1 - p_1(t))^{1 - w_{ij}(t)}, & \{i, j\} \cap \cI_\new(t) \neq \emptyset,
    \end{cases}\label{eqn:proxy_network}
\end{align}
where $p_{1|0}(t) = (1 - \phi)p_1(t)$ and $p_{1|1}(t) = 1 - (1 - \phi)\left(1 - p_1(t)\right)$.

The first two cases in \eqref{eqn:proxy_network} assign the same conditional probability we would expect from the complete data model for connections among returners, and the third case covers connections for which at least one observed individual is a newcomer. Because no previous connection information is available for newcomers, the probability of a connection is assigned the stationary density of the network, just as it is for connections at $t = 1$ in the complete data model. When at least one of $i$ or $j$ is not in $\cI_\obs(t)$, the connection is simply ignored. Finally, define the probability mass functions for connections across observed individuals $i, j \in \cI_\obs(1)$ at time $t = 1$ in the same way as the complete data model as $\tilde{p}\left(w_{ij}(1)\right) = p_1(1)^{w_{ij}(1)}(1 - p_1(1))^{1 - w_{ij}(1)}$. We implement Gibbs updates for the non-ignored connections using the proxy conditional probabilities in lieu of the complete probabilities. 

We define the proxy probability mass function for all relevant social connections across at all time points, $\mathbf{W}_\obs$, as
\begin{align}
  \tilde{p}(\mathbf{W}_\obs | \boldsymbol\delta_p, \phi) =
    \prod_{t = 1}^T \prod_{i < j \in \cI_\obs(t)}^J \tilde{p}\left(w_{ij}(t) | w_{ij}(t - 1)\right),
\end{align}
where $\tilde{p}\left(w_{ij}(1) | w_{ij}(0)\right)$ is understood to mean $\tilde{p}\left(w_{ij}(1)\right)$. The proxy probability mass function can be used to compute the full-conditional distributions required for Metropolis updates of $\boldsymbol\delta_p$ and $\phi$ within a comprehensive MCMC algorithm.

Because individuals receive multiple labels throughout the study period, the pairwise connections $w_{ij}(t)$ are not of primary interest for our drone-based movement data because they are only relevant for particular, uninterrupted sequences of observations. However, if it can be assumed that the censoring process is independent of individuals' positions in the social network, it is natural to think that the subset of network connections at each time might be representative of the entire social network and thus global statistics computed for the sub-network will constitute unbiased estimates for the complete network. Further, the density, $p_1(t)$ and stability, $\phi$ of the sub-network are informed by the number and durability of connections at each time, and are independent of the node labels associated with each potential connection. Thus, the censored, multiply-labeled data should contain information about two key aspects of social behavior: the time-varying density and stability of the network. Indeed, past research has shown that some global network features are reliably estimated from networks built from subset of nodes \citep{kossinets2006effects}.

\section{Simulation study} \label{sec:sim}
To assess the validity of inference derived from our proposed proxy likelihood, we conducted an extensive simulation study. We defined the vector of covariates at each time, $\mathbf{x}(t) = \left(x_1(t), x_2(t), x_3(t)\right)^\prime$, the same way for each of the GLMs for alignment, attraction, and network density ($\alpha(t), \beta(t), p_1(t)$, respectively). Elements in the length-three vector include an intercept, $x_1(t) = 1$, and two additional variables indicating whether $t$ corresponded to a time during, $x_2(t)$, or after, $x_3(t)$, exposure. Thus, the second two regression coefficients in each GLM describe the effect of active and recently terminated sonar exposure on the dynamic characteristics of social movement relative to an overall baseline. For the purpose of our simulation study, we used the same covariate construction and assumed the study period was equally divided into before, during, and after exposure intervals. 

We simulated trajectories for five interacting individuals over 300 time steps according to the social movement model for a broad combination of parameter values (512 unique combinations; Table~\ref{tab:simulation_combos}). For each combination, we generated 9 censored data sets according to different censoring schemes (see Section~\ref{sec:censoring}). Finally, for each combination of model parameters and censoring scheme, we simulated 12 sets of trajectories, for a total of 55,296 unique sets of trajectories. We obtained samples from posterior distributions based on the complete data sets using the exact likelihood, and samples from each of the 9 censored data sets using the proxy likelihood using the R package NIMBLE \citep{nimble2017, nimble2022}. We then examined the marginal posterior distributions of all model parameters. Code used to carry out the simulation study is available as part of the Supplementary Materials. 

\subsection{Simulated censoring}\label{sec:censoring}
To simulate patterns of censoring similar to what we observe in the data for our motivating application, we used the following procedure. Each individual was assigned to an initial ($t = 1$) observed or missing state with probability 0.5. Independently for each individual, a sequence of entry and exit times were sampled such that the duration of the observation or missingness is Poisson-distributed with parameters $\lambda_\obs$, $\lambda_\miss$, respectively. The initial state and entry/exit times were used to determine when an individual is observed or missing under the realized censoring pattern, and multi-labeling occurred in the same way it takes place in the application to dolphin movement: whenever an individual appeared (at an entry time), it was assigned a new label. Thus, the greater the number of entry times, the larger the number of unique individual identifiers for a fixed true population size. For each collection of simulated, uncensored trajectories, we considered nine possible censoring schemes according to all combinations of $\lambda_\obs, \lambda_\miss \in \{5, 10, 20\}$. The 3 parameter values used in the censoring process correspond to the expected durations of observation and missingness. Thus, qualitatively, our 9 sets of parameter combinations capture a variety of scenarios combining short, medium, and long periods of observation and missingness.

\begin{table}
\caption{Parameter values and censoring schemes used in simulation.}
\begin{tabular}{ll}
\hline \hline
parameter & space \\
\hline
$\boldsymbol\delta_\alpha, \boldsymbol\delta_\beta, \boldsymbol\delta_p$ & 
  $\{-2, 2\} \times \{-1, 1\} \times \{-0.5, 0.5\}$ \\
$\sigma^2$ & 1  \\
$\phi$ & $1 / (1 + e^{-2}) \approx 0.88$ \\
$\lambda_\mathrm{obs}, \lambda_\mathrm{miss}$ & $\{5, 10, 20\}$ \\
\hline
\end{tabular}
\label{tab:simulation_combos}\\
\vspace{1em}
\footnotesize{\textit{All possible combinations of these values (i.e., 2 values for each of 3 vectors of 3 regression coefficients each) were used for a total of $2^3 \times 2^3 \times 2^3 = 512$ unique combinations of coefficients. Each coefficient combination was used to simulate 10 realizations of trajectories, and $2^2$ unique censoring schemes were applied to each realization.}}
\end{table}

Our primary interest was in determining whether there was evidence of any systematic location shifts between the complete and proxy-based posteriors, as such a shift would suggest inferential biases that would render inference untrustworthy. Thus, for each model parameter and repetition in the simulation study, we computed the difference between the medians of samples from the marginal posterior distributions using the complete and proxy likelihoods as $d_\theta = \mathrm{med}_{L}\left(\theta|\bmu\right) - \mathrm{med}_{\tilde{L}}\left(\theta|\bmu_\obs\right)$. Evidence for systematic location bias in the posterior distributions based on proxy likelihoods was therefore observable as values of $d_\theta$ consistently above or below 0 across the 12 repetitions. 

\begin{figure}[ht]
\begin{center}
\hspace*{-0.12\textwidth}
\includegraphics[width = 1.24\textwidth]{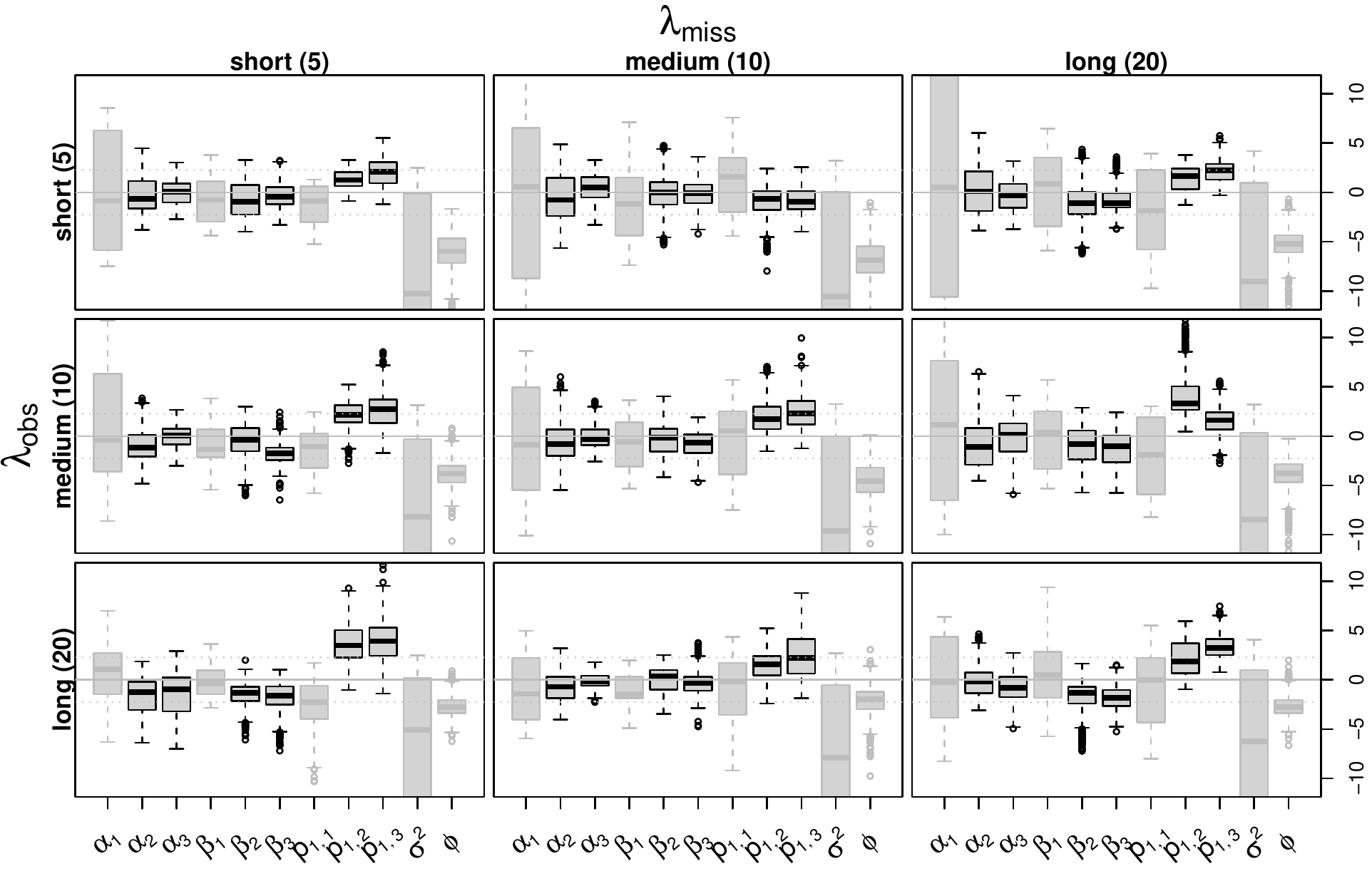}
\end{center}
\caption{Boxplots of standardized differences $\left(\frac{d_\theta - \overline{d_\theta}}{\mathrm{sd}(d_\theta) / \sqrt{10}}\right)$ across each set of ten repeated simulations with identical parameter configurations. Each of the four plots show differences for a single censoring scheme. Labels along the horizontal axis correspond to regression coefficients in the GLMs for alignment, $\alpha(t)$, attraction, $\beta(t)$, and network density, $p_1(t)$. Horizontal dashed lines show boundaries for statistical significance for a two-sided t-test at level 0.05.}
\label{fig:boxplots}
\end{figure}

Figure~\ref{fig:boxplots} shows boxplots of standardized differences $\left(\frac{d_\theta - \overline{d_\theta}}{\mathrm{sd}(d_\theta) / \sqrt{12}}\right)$ for all 11 parameters across each set of 12 repeated simulations with identical parameter configurations. Black outlines highlight the parameters of relevance in our application. Parameters with the most concerning bias across all four censoring schemes are $\delta_{p,1,2}$, $\delta_{p,1,3}$, $\sigma^2$, and $\phi$, the latter two of which are nuisance parameters with respect to our scientific goals of determining the effects of sonar on social behavior.

Our simulation study sheds light on which model parameters are potentially estimable from drone-based data using the proposed proxy likelihood. Some parameters display substantial bias. For a few parameters, a closer examination of the proxy likelihood construction suggests a qualitative explanation. Heuristically, $\tilde{L}(t, \bmu_\ret(t))$ contains information from the data about connections among the $n_\ret(t)$ individuals via those individuals' tendencies to exhibit attraction and alignment with each other. Social connections involving unobserved individuals are ignored, and implicitly taken to be absent, rather than marginalized over. Ignoring connections involving unobserved individuals means that we will generally infer fewer neighbors for observed individuals than we would if we observed all individuals \citep{kossinets2006effects}. Thus, the inferred ego-network sizes, $w_{i+}(t)$, will be biased downward, which in turn introduces a bias in the estimated posterior distributions for the intercept for the generalized linear model on the stationary density of the network, and $\sigma^2$. To see how the bias in $\sigma^2$ arises, recall that the marginal precisions of the multivariate distribution for $\bmu_i(t)$ are equal to $\sigma^{-2}\sum_{i = 1}^n w_{ij}(t)$. The sum over $w_{ij}(t)$ will be underestimated, and to compensate, the posterior distribution of $\sigma^2$ will shift toward 0 so that the marginal variances are consistent with the observed step sizes in the data.

We suspect that a related effect of ignoring connections between observed and unobserved individuals could explain the negative bias in $\phi$. As individuals leave the field of view, previously established connections will implicitly dissolve because, as mentioned above, the proxy likelihood assumes zero connections between observed and unobserved individuals. Thus, the overall rate of change for connections in the dynamic network will be artificially increased because some connections between observed and previously-observed individuals will in fact endure, in conflict with the assumptions underlying the proxy likelihood.

\section{Application} \label{sec:app}

We implemented our methodology on drone-measured movement data of Risso's dolphins (\textit{G. Griseus}) near Santa Catalina Island, off the coast of California. Experimental design, image collection, and spatial tracking of dolphins is detailed in \citet{durban2022integrating}. Observed locations were derived from images taken at one-second intervals for a total observation period of 26.5 minutes. An individual could be unambiguously tracked between images based on expected group spacing \cite[see][]{durban2022integrating} for a median of 26 seconds (min~=~2, 1st quartile~=~7, 3rd quartile~=~74, max~=~325). Individual tracks were curtailed when a dolphin submerged deep underwater, beyond view, or swam outside of the camera field of view. The same individuals returned seconds or minutes later, thus inducing the multi-labeling issue discussed in Section~\ref{sec:intro}. The total number of concurrently observed dolphins varied over the course of the study period from as few as 0 to as many as 9, and had a median value of 5 individuals (see Supplementary Materials for an animation of the data).

Data were gathered as part of a controlled exposure experiment to investigate Risso's dolphins' response to the presence of naval sonar. An attenuated sonar source \cite[see][]{durban2022integrating} was introduced at a distance of 1.5km from the dolphin group approximately 14 minutes after observations began. Sonar pings occurred every 25 seconds for approximately 10 minutes, after which observation of the dolphins continued for approximately 2.5 additional minutes. The primary scientific research goal of our analysis was to quantify evidence that sonar impacted the social behavior of the study group during and/or immediately following exposure.

\begin{figure}[ht]
\begin{center}
\includegraphics[width = \textwidth]{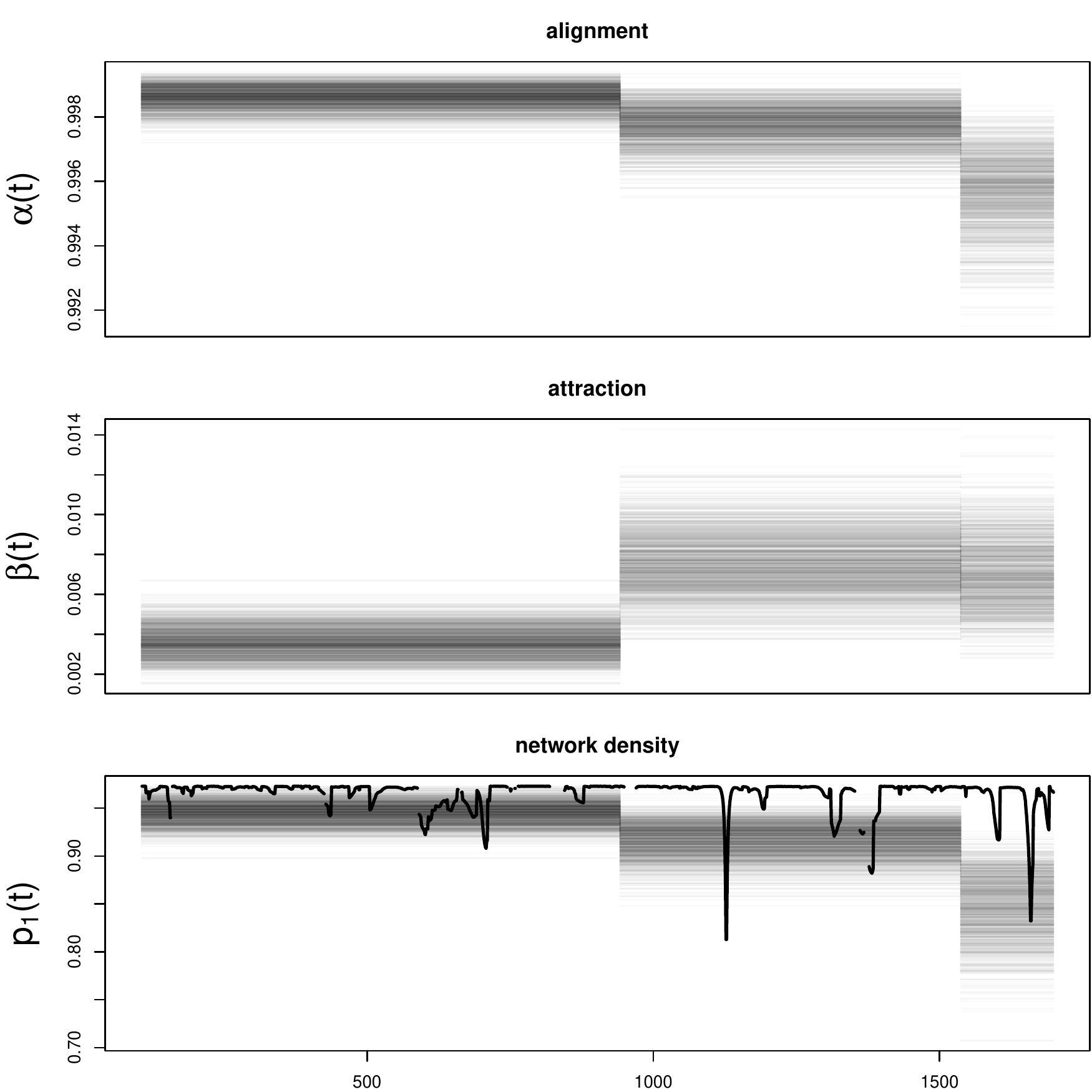}
\end{center}
\caption{Summaries of posterior distributions for $\alpha(t)$, $\beta(t)$, and $p_1(t)$. Left-most segments correspond to the phase of the CEE before the sonar begins, middle segments to the phase of the CEE while sonar is active, and right segments to the phase immediately following sonar exposure. The curve in the bottom plot shows the pointwise posterior mean of the density of connections among observed individuals.}
\label{fig:dynamic_behavior}
\end{figure}

We defined the vector of covariates at each time, $\mathbf{x}(t) = \left(x_1(t), x_2(t), x_3(t)\right)^\prime$, the same way for each of the GLMs for alignment, attraction, and network density ($\alpha(t), \beta(t), p_1(t)$, respectively), as we did in Section~\ref{sec:sim}. We fit the proposed model to the data within a Bayesian framework by specifying priors for all unknown variables, and we quantified evidence for changes in dolphin behavior during and after exposure to sonar by examining posterior distributions of the regression coefficients, $\boldsymbol\delta_\alpha$, $\boldsymbol\delta_\beta$, and $\boldsymbol\delta_p$. 
Figure~\ref{fig:dynamic_behavior} shows samples from the posterior distribution of each time-varying parameter in the social movement model. Each semi-transparent line segment corresponds to a sample from the posterior distribution, and the steps in the function occur at the changes between the before, during, and after phases of exposure. There is some evidence of small decreases in the alignment parameter at both the onset and conclusion of exposure (Figure~\ref{fig:dynamic_behavior}, top), an increase in the attraction parameter at the onset of exposure (Figure~\ref{fig:dynamic_behavior}, middle), and a decrease in network density (Figure~\ref{fig:dynamic_behavior}, bottom). The presence of sonar is associated with an increase in the attractive effect and a decrease in both the alignment effect and the probability of connection between dolphins. 

Table~\ref{tab:compare} provides posterior probabilities for an increase in each time-varying parameter between each pair of experiment phases. For example, attraction is has a high probability of being greater during both the exposure and post-exposure phases than before exposure (Table~\ref{tab:compare}, second row), while alignment is extremely unlikely to increase above the before-exposure level (first row). Thus, \textit{G. Griseus} individuals in this group of Risso's dolphins are more likely to exhibit an inclination towards the mean position of connected individuals and less likely to develop and maintain social connections when exposed to sonar. The equal-tailed 95\% credible interval for $\phi$ was $(0.935, 0.972)$, which corresponds to expected durations for social connections of between about 15 and 35 seconds; however, our simulation study suggests this interval may be biases strongly downward, implying more long-lasting real connections. 

Behaviors associated with the inferred changes in parameter values could be quite subtle, given the magnitudes seen in our application. Thus, it may be challenging to detect behavioral shifts simply by watching raw footage of the swimming dolphins. However, behavior consistent with similar shifts in parameter values, but larger magnituds, might, for example, entail dolphins shifting from a few large groups into several smaller, more tightly arranged groups upon exposure to sonar. 

\begin{table}[ht]
\caption{Posterior probabilities comparing changes in different characteristics of social behavior in response to sonar exposure.}
\begin{tabular}{llll}
\hline \hline
 & before$<$during & before$<$after & during$<$after \\ 
\hline
$\alpha(t)$ & 0.097 & 0.000 & 0.028 \\ 
$\beta(t)$ & 0.991 & 0.985 & 0.420 \\ 
$p_1(t)$ & 0.048 & 0.000 & 0.026 \\
\hline
\end{tabular}
\label{tab:compare}\\
\vspace{1em}
\footnotesize{\textit{Rows correspond to behavior characteristics (alignment, attraction, network density), and columns correspond to comparisons among the three different experimental intervals (before, during, after).}}
\end{table}

To attempt to match our application with one of the censoring schemes explored in the simulation study, we derived estimators for $\lambda_\obs$ and $\lambda_\miss$ based on the observed arrival and departure times and the true total population, $J$ (see Supplementary Materials). The estimators use the sample mean of the durations of consecutive observations as an estimate of $\lambda_\obs$, and leverage the relationship between $\lambda_\obs$ and $\lambda_\miss$, and the marginal expected proportion of observed individuals in each frame to derive an estimate of $\lambda_\miss$ for situations where an estimate of the total population size, $J$, is available. Based on the expert opinion of the team responsible for gathering the application data, we estimate that approximately 17 dolphins were present in the study population. After scaling our estimates of $\lambda_\obs$ and $\lambda_\miss$ to reflect the difference in number of time steps between the simulation study and application, we find our application is equivlant to simulation study values of $\lambda_\obs \approx 9$ and $\lambda_\miss \approx 20$, or a ``medium-long'' scenario (second row, third column of Figure~\ref{fig:boxplots}).

According to our simulation study, the most probable biases present in our marginal posterior approximations would be in the effects of sonar on network density, which would be positive (see Figure~\ref{fig:boxplots}). Thus, the small negative shifts in $p_1(t)$ shown in the bottom plot of Figure~\ref{fig:dynamic_behavior} may be conservative. That is, the effect of sonar and post-sonar exposure on reducing network density may be underestimated as a result of our implementation approach. In addition, our simulation study indicated the presence of substantial systematic bias in the negative direction for $\phi$, suggesting that social connections may be more durable than the posterior suggests.

Although particular pairwise network connections among the multiply-labeled individuals at each time $t$ have little scientific meaning, global network characteristics such as mean degree, network density, etc. can offer valuable insight about characteristics of the entire population. As one example, the pointwise posterior mean of the normalized mean degree for the dynamic social network is shown as a black line in the bottom plot of Figure~\ref{fig:dynamic_behavior}. Consistent with the posterior distribution of $p_1(t)$, the mean number of connections reaches its lowest values during and following sonar exposure. 

\section{Discussion} \label{sec:dis}

There is an increasing awareness of the impact of social connections' on the behavior of a population. Individuals have the potential to influence and be influenced by the movement of the collective population \citep{torney2018inferring, calabrese2018disentangling}, which can be captured in a socially informed movement model \citep{haydon2008socially}. One way to interpret these influences is to define social interactions through an attraction mechanism, in which individuals move towards those to which they are connected \citep{couzin2005effective, delgado2014statistical, scharf2016dynamic}. Other interpretations include modeling social interactions related to a shared moving target \citep{langrock2014modelling, niu2016modeling} or within the social hierarchies of a population \citep{milner2021modelling}.

Statistical models for movement have only very recently begun to model the positions of potentially interacting individuals jointly, thereby explicilty acknowledging the role social interactions can play in animal movement. Many recently-developed statistical models represent extensions of the individual-based movement models first proposed many decades ago. For instance, \cite{scharf2016dynamic} is essentially a multivariate generalization of a discrete-time Gaussian random walk with attraction \citep[e.g.,][]{hooten_discrete-time_2017}, and \cite{niu2016modeling} is a multivariate extensions of the continuous-time Ornstein-Uhlenbeck model \citep{blackwell_bayesian_2003, johnson_continuous-time_2008, hooten_continuous-time_2017}. Historically, point-process models have also been used to make inference about behavior and space-use from movement data, but we are unaware of any recent methods for the joint analysis of multiple trajectories explicitly motivated by this perspective.

We extended an existing Bayesian hierarchical movement model by making use of a latent unlabeled social network. We developed a proxy for the exact likelihood that is intractable due to the multiply-labeled nature of our data, which we used to study global network characteristics providing behavioral inference for a population. Although our model allows for the inference of ecologically relevant characteristics of the global social network, it does not allow inference of pairwise connections between specific individuals. The lack of information at the individual level may be a limitation for researchers interested in estimates of pairwise social connections; however, information on the behavioral characteristics of an entire population (a dolphin social group, in the case of our example application) can still provide valuable scientific insight.

In addition to enabling drone-measured movement data as a mechanism for analyzing animal behavior and determining the effect of human encroachment on animal habitats, our framework can be generalized to other forms of movement data with partially observed individuals. Potential extensions are not limited to animal populations. Our model may also be used to analyze the collective movement of humans while protecting privacy at the individual level. 

A notable challenge of our methodology is computational expense. Our application study analyzing a small group of individuals with at most 9, and on average 5, concurrently-observed individuals had a run-time of about two hours and required about 15GB of memory. Based on additional applications undertaken by the authors, we estimate that the current implementation of our proposed method using NIMBLE will experience memory-related bottlenecks that prevent application to populations with above on average 10 concurrently-observed individuals over a similar number of time steps. However, an increase in computational burden may be avoided by studying subsets of a population. Such an extension is justifiable as our implementation and simulation studies reveal an entire study population does not have to be visible for the duration of the study period in order to make inference about the network and some of its characteristics.

Another limitation to our proposed method is the consistent bias the proxy likelihood induces in some model parameters, as we discuss in Section~\ref{sec:sim}. We note that many of the empirically-observed issues related to bias may result from underestimating the number and duration of connections in the social network, in particular between observed and unobserved individuals. This proposed explanation suggests some possible directions for remediation, and we have indeed explored some possible adjustments to the proxy likelihood intended to correct under-counting connections. However, we have not yet identified an alternative proxy likelihood with consistently better properties, although this remains a focus for future research.

While future studies can be used for further validation purposes and to determine whether analyzing larger groupings and/or populations is feasible, our methodology provides researchers with an outlet for analyzing previously inaccessible movement data while inferring behavioral characteristics of a population on a global level. 

\section*{Data Availability}

The authors declare that the data supporting the findings of this study are available within the article's supplementary information files. \textit{Supplementary files containing all data and code needed to reproduce the figures and results in this manuscript will be provided upon acceptance for publication.}

\begin{acks}[Acknowledgments]
Funding for the collection of the dolphin data was provided by the U.S. Navy’s Office of Naval Research (Awards N000141713132, N0001418IP-00021, N000141712887, N000141912572). Drone flights over dolphins were authorized by research permit 19091 from the National Marine Fisheries Service (NMFS), and the controlled sonar exposure experiment was conducted under NMFS permit 19116.
\end{acks}


\bibliographystyle{ba}
\bibliography{drone}

\begin{thebibliography}{30}
\newcommand{\enquote}[1]{``#1''}
\expandafter\ifx\csname natexlab\endcsname\relax\def\natexlab#1{#1}\fi
\expandafter\ifx\csname url\endcsname\relax
  \def\url#1{{\tt #1}}\fi
\expandafter\ifx\csname urlprefix\endcsname\relax\def\urlprefix{URL }\fi
\ifx\endbibitem\undefined \let\endbibitem\relax\fi

\bibitem[{Blackwell(2003)}]{blackwell_bayesian_2003}
Blackwell, P.~G. (2003).
\newblock \enquote{Bayesian inference for {Markov} processes with diffusion and
  discrete components.}
\newblock {\em Biometrika\/}, 90(3): 613--627.
\endbibitem

\bibitem[{Bode et~al.(2012)Bode, Franks, Wood, Piercy, Croft, and
  Codling}]{bode2012distinguishing}
Bode, N.~W., Franks, D.~W., Wood, A.~J., Piercy, J.~J., Croft, D.~P., and
  Codling, E.~A. (2012).
\newblock \enquote{Distinguishing social from nonsocial navigation in moving
  animal groups.}
\newblock {\em The American Naturalist\/}, 179(5): 621--632.
\endbibitem

\bibitem[{Calabrese et~al.(2018)Calabrese, Fleming, Fagan, Rimmler, Kaczensky,
  Bewick, Leimgruber, and Mueller}]{calabrese2018disentangling}
Calabrese, J.~M., Fleming, C.~H., Fagan, W.~F., Rimmler, M., Kaczensky, P.,
  Bewick, S., Leimgruber, P., and Mueller, T. (2018).
\newblock \enquote{Disentangling social interactions and environmental drivers
  in multi-individual wildlife tracking data.}
\newblock {\em Philosophical Transactions of the Royal Society B: Biological
  Sciences\/}, 373(1746): 20170007.
\endbibitem

\bibitem[{Carretta et~al.(2019)Carretta, Forney, Oleson, Weller, Lang, Baker,
  Muto, Hanson, Orr, Huber, Lowry, Barlow, Moor, Lynch, Carswell, and
  Jr.}]{caretta2019stock}
Carretta, J.~V., Forney, K.~A., Oleson, E.~M., Weller, D.~W., Lang, A.~R.,
  Baker, J., Muto, M.~M., Hanson, B., Orr, A.~J., Huber, H., Lowry, M.~S.,
  Barlow, J., Moor, J.~E., Lynch, D., Carswell, L., and Jr., R. L.~B. (2019).
\newblock \enquote{U.S. Pacific Marine Mammal Stock Assessments: 2018.}
\newblock Technical report, Southwest Fisheries Science Center (U.S.).
\endbibitem

\bibitem[{Couzin et~al.(2005)Couzin, Krause, Franks, and
  Levin}]{couzin2005effective}
Couzin, I.~D., Krause, J., Franks, N.~R., and Levin, S.~A. (2005).
\newblock \enquote{Effective leadership and decision-making in animal groups on
  the move.}
\newblock {\em Nature\/}, 433(7025): 513--516.
\endbibitem

\bibitem[{Dawson et~al.(2017)Dawson, Bowman, Leunissen, and
  Sirguey}]{dawsonphotgrammetry2017}
Dawson, S., Bowman, M., Leunissen, E., and Sirguey, P. (2017).
\newblock \enquote{Inexpensive Aerial Photogrammetry for Studies of Whales and
  Large Marine Animals.}
\newblock {\em Frontiers in Marine Science\/}, 4: 366.
\endbibitem

\bibitem[{{de Valpine} et~al.(2022){de Valpine}, Paciorek, Turek, Michaud,
  Anderson-Bergman, Obermeyer, {Wehrhahn Cortes}, Rodrìguez, {Temple Lang},
  and Paganin}]{nimble2022}
{de Valpine}, P., Paciorek, C., Turek, D., Michaud, N., Anderson-Bergman, C.,
  Obermeyer, F., {Wehrhahn Cortes}, C., Rodrìguez, A., {Temple Lang}, D., and
  Paganin, S. (2022).
\newblock {\em {NIMBLE}: {MCMC}, Particle Filtering, and Programmable
  Hierarchical Modeling\/}.
\newblock {R} package version 0.12.2.
\endbibitem

\bibitem[{{de Valpine} et~al.(2017){de Valpine}, Turek, Paciorek,
  Anderson-Bergman, {Temple Lang}, and Bodik}]{nimble2017}
{de Valpine}, P., Turek, D., Paciorek, C., Anderson-Bergman, C., {Temple Lang},
  D., and Bodik, R. (2017).
\newblock \enquote{Programming with models: writing statistical algorithms for
  general model structures with {NIMBLE}.}
\newblock {\em Journal of Computational and Graphical Statistics\/}, 26:
  403--413.
\endbibitem

\bibitem[{Delgado et~al.(2014)Delgado, Penteriani, Morales, Gurarie, and
  Ovaskainen}]{delgado2014statistical}
Delgado, M. d.~M., Penteriani, V., Morales, J.~M., Gurarie, E., and Ovaskainen,
  O. (2014).
\newblock \enquote{A statistical framework for inferring the influence of
  conspecifics on movement behaviour.}
\newblock {\em Methods in Ecology and Evolution\/}, 5(2): 183--189.
\endbibitem

\bibitem[{Durban et~al.(2015)Durban, Fearnbach, Barrett-Lennard, Perryman, and
  Leroi}]{durban2015hexacopter}
Durban, J., Fearnbach, H., Barrett-Lennard, L., Perryman, W., and Leroi, D.
  (2015).
\newblock \enquote{Photogrammetry of killer whales using a small hexacopter
  launched at sea.}
\newblock {\em Journal of Unmanned Vehicle Systems\/}, 3.
\endbibitem

\bibitem[{Durban et~al.(2022)Durban, Southall, Calambokidis, Casey, Fearnbach,
  Joyce, Fahlbusch, Oudejans, Fregosi, Friedlaender, Kellar, and
  Visser}]{durban2022integrating}
Durban, J.~W., Southall, B., Calambokidis, J., Casey, C., Fearnbach, H., Joyce,
  T.~W., Fahlbusch, J., Oudejans, M.~G., Fregosi, S., Friedlaender, A.~S.,
  Kellar, N.~M., and Visser, F. (2022).
\newblock \enquote{Integrating remote sensing methods during controlled
  exposure experiments to quantify group responses of dolphins to navy sonar.}
\newblock {\em Marine Pollution Bulletin\/}, 174.
\endbibitem

\bibitem[{Hartman et~al.(2008)Hartman, Visser, and
  Hendriks}]{hartman2008stratified}
Hartman, K.~L., Visser, F., and Hendriks, A.~J. (2008).
\newblock \enquote{Social structure of Risso’s dolphins (Grampus griseus) at
  the Azores: a stratified community based on highly associated social units.}
\newblock {\em Canadian Journal of Zoology\/}, 86(4): 294--306.
\endbibitem

\bibitem[{Haydon et~al.(2008)Haydon, Morales, Yott, Jenkins, Rosatte, and
  Fryxell}]{haydon2008socially}
Haydon, D.~T., Morales, J.~M., Yott, A., Jenkins, D.~A., Rosatte, R., and
  Fryxell, J.~M. (2008).
\newblock \enquote{Socially informed random walks: incorporating group dynamics
  into models of population spread and growth.}
\newblock {\em Proceedings of the Royal Society B: Biological Sciences\/},
  275(1638): 1101--1109.
\endbibitem

\bibitem[{Hooten et~al.(2017{\natexlab{a}})Hooten, Johnson, McClintock, and
  Morales}]{hooten_continuous-time_2017}
Hooten, M.~B., Johnson, D.~S., McClintock, B.~T., and Morales, J.~M.
  (2017{\natexlab{a}}).
\newblock \enquote{Continuous-{Time} {Models}.}
\newblock In {\em Animal {Movement}: {Statistical} {Models} for {Telemetry}
  {Data}\/}. CRC Press.
\endbibitem

\bibitem[{Hooten et~al.(2017{\natexlab{b}})Hooten, Johnson, McClintock, and
  Morales}]{hooten_discrete-time_2017}
--- (2017{\natexlab{b}}).
\newblock \enquote{Discrete-{Time} {Models}.}
\newblock In {\em Animal {Movement}: {Statistical} {Models} for {Telemetry}
  {Data}\/}. CRC Press.
\endbibitem

\bibitem[{Johnson et~al.(2008)Johnson, London, Lea, and
  Durban}]{johnson_continuous-time_2008}
Johnson, D.~S., London, J.~M., Lea, M.-A., and Durban, J.~W. (2008).
\newblock \enquote{Continuous-time correlated random walk model for animal
  telemetry data.}
\newblock {\em Ecology\/}, 89(5): 1208--1215.
\endbibitem

\bibitem[{Kossinets(2006)}]{kossinets2006effects}
Kossinets, G. (2006).
\newblock \enquote{Effects of missing data in social networks.}
\newblock {\em Social Networks\/}, 28(3): 247--268.
\endbibitem

\bibitem[{Langrock et~al.(2014)Langrock, Hopcraft, Blackwell, Goodall, King,
  Niu, Patterson, Pedersen, Skarin, and Schick}]{langrock2014modelling}
Langrock, R., Hopcraft, J. G.~C., Blackwell, P.~G., Goodall, V., King, R., Niu,
  M., Patterson, T.~A., Pedersen, M.~W., Skarin, A., and Schick, R.~S. (2014).
\newblock \enquote{Modelling group dynamic animal movement.}
\newblock {\em Methods in Ecology and Evolution\/}, 5(2): 190--199.
\endbibitem

\bibitem[{Lusseau~David(2004)}]{lusseau2004role}
Lusseau~David, N. M.~E. (2004).
\newblock \enquote{Identifying the role that animals play in their social
  networks.}
\newblock {\em Proc R Soc Lond B\/}, (271): 477--481.
\endbibitem

\bibitem[{Milner et~al.(2021)Milner, Blackwell, and Niu}]{milner2021modelling}
Milner, J.~E., Blackwell, P.~G., and Niu, M. (2021).
\newblock \enquote{Modelling and inference for the movement of interacting
  animals.}
\newblock {\em Methods in Ecology and Evolution\/}, 12(1): 54--69.
\endbibitem

\bibitem[{Niu et~al.(2016)Niu, Blackwell, and Skarin}]{niu2016modeling}
Niu, M., Blackwell, P.~G., and Skarin, A. (2016).
\newblock \enquote{Modeling interdependent animal movement in continuous time.}
\newblock {\em Biometrics\/}, 72(2): 315--324.
\endbibitem

\bibitem[{Niu et~al.(2020)Niu, Frost, Milner, Skarin, and
  Blackwell}]{niu2020modelling}
Niu, M., Frost, F., Milner, J.~E., Skarin, A., and Blackwell, P.~G. (2020).
\newblock \enquote{Modelling group movement with behaviour switching in
  continuous time.}
\newblock {\em Biometrics\/}, 78(1): 286--299.
\endbibitem

\bibitem[{Rice et~al.(2020)Rice, Rafter, Trickey, Wiggins, Baumann-Pickering,
  and Hildebrand}]{rice2019acoustic}
Rice, A.~C., Rafter, M., Trickey, J.~S., Wiggins, S.~M., Baumann-Pickering, S.,
  and Hildebrand, J.~A. (2020).
\newblock \enquote{Passive Acoustic Monitoring for Marine Mammals in the SOCAL
  Range Complex July 2018-May 2019.}
\newblock Technical report, Scripps Institution of Oceanography.
\endbibitem

\bibitem[{Rue and Held(2005)}]{rue2005gaussian}
Rue, H. and Held, L. (2005).
\newblock {\em Gaussian Markov Random Fields: Theory and Applications\/}.
\newblock CRC press.
\endbibitem

\bibitem[{Scharf and Buderman(2020)}]{scharf2020animal}
Scharf, H.~R. and Buderman, F.~E. (2020).
\newblock \enquote{Animal movement models for multiple individuals.}
\newblock {\em Wiley Interdisciplinary Reviews: Computational Statistics\/},
  12(6): e1506.
\endbibitem

\bibitem[{Scharf et~al.(2016)Scharf, Hooten, Fosdick, Johnson, London, and
  Durban}]{scharf2016dynamic}
Scharf, H.~R., Hooten, M.~B., Fosdick, B.~K., Johnson, D.~S., London, J.~M.,
  and Durban, J.~W. (2016).
\newblock \enquote{Dynamic social networks based on movement.}
\newblock {\em The Annals of Applied Statistics\/}, 10(4): 2182--2202.
\endbibitem

\bibitem[{Scharf et~al.(2018)Scharf, Hooten, Johnson, and
  Durban}]{scharf2018processes}
Scharf, H.~R., Hooten, M.~B., Johnson, D.~S., and Durban, J.~W. (2018).
\newblock \enquote{Process convolution approaches for modeling interacting
  trajectories.}
\newblock {\em Environmetrics\/}, 29(3).
\endbibitem

\bibitem[{Torney et~al.(2018)Torney, Lamont, Debell, Angohiatok, Leclerc, and
  Berdahl}]{torney2018inferring}
Torney, C.~J., Lamont, M., Debell, L., Angohiatok, R.~J., Leclerc, L.-M., and
  Berdahl, A.~M. (2018).
\newblock \enquote{Inferring the rules of social interaction in migrating
  caribou.}
\newblock {\em Philosophical Transactions of the Royal Society B: Biological
  Sciences\/}, 373(1746): 20170385.
\endbibitem

\bibitem[{Weiss et~al.(2021)Weiss, Ellis, and Croft}]{weiss2021diversity}
Weiss, M.~N., Ellis, S., and Croft, D.~P. (2021).
\newblock \enquote{Diversity and Consequences of Social Network Structure in
  Toothed Whales.}
\newblock {\em Canadian Journal of Zoology\/}, 921.
\endbibitem

\bibitem[{Whitehead and Rendell(2021)}]{whitehead2021cultural}
Whitehead, H. and Rendell, L. (2021).
\newblock {\em The Cultural Lives of Whales and Dolphins\/}.
\newblock University of Chicago Press.
\endbibitem

\end{thebibliography}

\end{document}